\documentclass[12pt,a4paper]{article}
\newcommand{\en}{\end{eqnarray}}
\newcommand{\bea}{\begin{eqnarray}}
\newcommand{\eea}{\end{eqnarray}}

\newcommand{\be}{\begin{equation}}
\newcommand{\ee}{\end{equation}}
\newcommand{\co}{\; \; ,}

\newcommand{\scs}{\co \;}
\newcommand{\sem}{ \; \; ; \;}
\newcommand{\per}{ \; .}

\newcommand{\GeV}{\mbox{GeV}}
\newcommand{\MeV}{\, \mbox{MeV}}

\begin{document}
\thispagestyle{empty}
\begin{flushright}
\small{BUTP-2000/3\\
HIP-2000-02/TH}\\[1cm]
\end{flushright}
\vskip1cm
\begin{center}

\renewcommand{\thefootnote}{\fnsymbol{footnote}}
{\bf{\large{
SIGMA-TERM PHYSICS}}}\\[1cm]

J\"urg Gasser\footnote[2]{Invited talk given  at the Workshop
{\em Physics and Detectors for DA$\Phi$NE},
 Frascati, Nov. 16-19, 1999. To appear in the Proceedings.}$^1$
     and Mikko E. Sainio$^2$\\[1cm]
 $^1$Institut f\"ur  Theoretische Physik, Universit\"at  Bern,\\
Sidlerstrasse 5, CH-3012 Bern, Schweiz\\
e-mail: gasser@itp.unibe.ch\\[.5cm]
$^2${ Helsinki Institute of Physics,
 Department of Physics,}\\ {University of Helsinki,
P.O.B. 9, FIN-00014 Helsinki, Finland}\\
e-mail: sainio@rock.helsinki.fi\\[1cm]

February 2000\\[1cm]

\end{center}
{\bf{Pacs:}} 11.30.Rd, 12.39.Fe, 13.75.Gx, 13.75.Jz, 13.75.Lb\\[.5cm]

\begin{center}
{\bf {Abstract}}
\end{center}
\noindent
We consider sigma-terms in $\pi\pi$, $\pi N$ and $K N$ scattering.
While the state of the art in $\pi\pi$ is in principle crystal clear,
 the kaon-nucleon case is largely unexplored. The pion-nucleon system is
in-between. We propose a list of topics to be
investigated in kaon-nucleon scattering, in order to make optimal use
of the precision data expected from DEAR.

\vskip1cm

\setcounter{equation}{0}
\section{Introduction}
The sigma-terms considered here are proportional to
 the following matrix elements of scalar quark currents in the
 framework of QCD,
\bea
\langle A|m_q\bar{q}q|A\rangle\sem q=u,d,s\sem A=\pi,K,N\per\nonumber
\eea
These matrix elements are of interest, because they are related
\begin{itemize}
\item
to the mass spectrum,
\item
to scattering amplitudes through Ward identities,
\item
to the  strangeness content of $A$,
\item
to quark mass ratios.
\end{itemize}
One may consider  sigma-terms and the strangeness content e.g. in

\vskip.5cm

\begin{tabular}{lll}
$\pi\pi\rightarrow\pi\pi$ &where&nearly everything is known,\\
$\pi N\rightarrow \pi N$& &much is known,\\
$\left.\hspace{-.2cm}\begin{array}{ll}
KN\rightarrow KN\\
\bar{K}N\rightarrow \bar{K}N\\
\pi K\rightarrow \pi K
\end{array}\right\}$&&little is known.
\end{tabular}

\vskip.5cm

In the following, we first illustrate the method to determine the
sigma-term in the $\pi\pi$ sector, where plenty of information is
available:
\begin{itemize}
\item[-]
from chiral perturbation theory \cite{weinberg,glan}(ChPT): the amplitude to
one loop  off-shell \cite{glan}
and  to two loops on-shell \cite{pipi6},
\item[-]the $\sigma$-term to two loops \cite{sigma6},
\item[-]the scattering amplitude from data and
        from Roy equations \cite{colangelo}.
\end{itemize}
Then, we briefly discuss the status in $\pi N$ and compare these
processes with kaon-nucleon scattering. We work in the isospin
symmetry limit $m_u=m_d$ throughout.

\setcounter{equation}{0}
\section{Sigma-term in $\pi\pi$ scattering}
We consider the elastic scattering process
\bea\label{eq21}
\pi^-(q)\pi^0(p)\rightarrow \pi^-(q')\pi^0(p')
\eea
in QCD.
We further introduce the standard Mandelstam variables
$s=(p+q)^2,t=(q'-q)^2, u=(p'-q)^2$, use $\nu=s-u$ and denote the
amplitude for the process (\ref{eq21}) by $A(t,\nu)$.
\subsection{The low-energy theorem for $\pi\pi$ scattering}
The low-energy theorem for the on-shell amplitude reads
\bea\label{eq22}
F_\pi^2A(t,\nu)={\Gamma_\pi(t)}+q'^\mu q^\nu r_{\mu\nu}\co
\eea
where  $\Gamma_\pi$ denotes the scalar form factor of the
pion,
\bea\label{eq23}
\Gamma_\pi(t)=\langle\pi^0(p')|\hat{m}(\bar{u}u
+\bar{d}d)|\pi^0(p)\rangle\sem \hat{m}=\frac{1}{2}(m_u+m_d)\per
\eea
 The quantity $r_{\mu\nu}$ is
not specified - the content of the theorem is the
statement that $r_{\mu\nu}$ has the same analytic properties as the
 scattering amplitude  itself - i.e., factoring the momenta $q,q'$ in the
manner shown in (\ref{eq22}) does not introduce kinematic
singularities in $r_{\mu\nu}$. Finally, the sigma-term  is
given by
\bea\label{eq24}
2M_\pi\sigma_\pi=\Gamma_\pi(0)\per
\eea
Therefore, in order to determine the sigma-term in this case, we need
to measure the scattering amplitude, calculate the remainder
$r_{\mu\nu}$, determine from this the scalar form factor
$\Gamma_\pi(t)$ using (\ref{eq22}), and evaluate it at $t=0$.
As is well known since the early days in sigma-term physics, one has
to make sure that the low-energy theorem (\ref{eq22}) is used in a
kinematic region where the remainder $r_{\mu\nu}$ is small -
otherwise, one introduces large uncertainties in the determination of
the sigma-term. To illustrate, we note that the relation
(\ref{eq22}) is true order by order in ChPT. At leading order, the
expressions read
\bea\label{eq25}
F_\pi^2A^{tree}&=&{t-M_\pi^2}\scs
\Gamma^{tree}_\pi={M_\pi^2}\scs
r_{\mu\nu}^{tree}=-{2g_{\mu\nu}}\per
\eea
Evaluating (\ref{eq22})  at  threshold, one finds that the remainder is
twice as big (and of opposite sign) as the contribution from the
scalar form factor itself,
whereas the scalar form factor is entirely given by the scattering
amplitude at $t=2M_\pi^2$,
\bea\label{eq26}
F_\pi^2A^{tree}&=&{{\Gamma_\pi^{tree}}}+
{\triangle_\pi^{tree}}\co\nonumber\\
-1&=&1 \hspace{.6cm}\,-2\hspace{.7cm}{\mbox{at threshold}}\co
\nonumber\\
 1&=&1 \hspace{.6cm}+\, 0\hspace{.7cm}{\mbox {at}}\,\, t=2M_\pi^2\co
\eea
where the numbers are given in pion mass units.
Therefore, it is not a good idea to use the low-energy theorem at
threshold, although it is perfectly valid there, of course.
 Beyond tree level, the advantageous region - where the remainder
is small - shrinks to
the Cheng-Dashen (CD) point
\bea\label{eq27}
t=2M_\pi^2\scs \nu=0\hspace{.5cm}{\mbox{Cheng-Dashen point}}\per
\eea
In the following, we write the relation (\ref{eq22})
 - evaluated at the CD point - in
the form
\bea\label{eq28}
F_\pi^2A^{CD}={\Gamma_\pi(2M_\pi^2)}+\triangle_\pi^{CD}\per
\eea
The chiral expansion neatly illustrates the advantage of this
choice. At the CD point, all quantities involved may be expanded in
powers of the pion mass. For the amplitude itself, this expansion
reads
\bea\label{eq29}
A^{CD}=\frac{M_\pi^2}{F_\pi^2}-\frac{13 M_\pi^4}{96\pi^2F_\pi^4}
\log{\frac{M_\pi^2}{\Lambda^2}}+O(M_\pi^6)\co
\eea
where the scale $\Lambda$ - which is independent of the pion mass - is
related to the low-energy constants ${l}_i$
in the chiral lagrangian \cite{glan},
\bea\label{eq210}
\Lambda=1.3\, \GeV\per
\eea
The logarithmic term in (\ref{eq29}) is an example of the infrared
singularities that show up in the chiral expansion \cite{glan}. The
main point is that the scalar form factor contains an analogous
singularity - they however nearly  cancel in the difference
$\triangle_\pi^{CD}$,
\bea\label{eq211}
\Gamma_\pi(2M_\pi^2)&=&{M_\pi^2}-\frac{3M_\pi^4}
{32\pi^2F_\pi^2}
\log{\frac{M_\pi^2}{\Lambda_1^2}}+O(M_\pi^6)\co\nonumber\\[.3cm]
\triangle_\pi^{CD}&=&-\frac{M_\pi^4}{24\pi^2F_\pi^2}
\log{\frac{M_\pi^2}{\Lambda_2^2}}+O(M_\pi^6)\co\nonumber\\[.3cm]
 \Lambda_1&=&1.1\, \GeV\scs
 \Lambda_2=1.8\, \GeV\per
\eea
Numerically, the relation (\ref{eq28}) becomes at one-loop accuracy
\bea\label{eq212}
{1.14}&=&\hspace{.5cm}1.09\hspace{.5cm}+ \hspace{.5cm}0.05
\nonumber\\
F_\pi^2A^{CD}&=&{\Gamma_\pi(2M_\pi^2)}\hspace{.2cm}
+\hspace{.4cm}\triangle_\pi^{CD}\co
\eea
where the numbers are again in pion mass units.
The remainder $\triangle_\pi^{CD}$ amounts to a correction of
$0.05M_\pi/2 \simeq 3.5\MeV$ -   the same size as in the
pion-nucleon case, see below.
\subsection{The sigma-term from data}
The above analysis and the relation (\ref{eq28}) indicate how the
sigma-term can be determined:

One evaluates
\begin{itemize}
\item
$A^{CD}$ from data on $\pi\pi\rightarrow\pi\pi$, using Roy
equations \cite{colangelo}.
\item
Dispersion relations for the form factor allow one to determine
\bea
\triangle_\sigma=\Gamma_\pi(2M_\pi^2)-\Gamma_\pi(0)\per
\eea
\item
The quantity $\triangle_\pi^{CD}$ is determined from ChPT.
\end{itemize}
Then, one has
\bea
2M_\pi\sigma_\pi=F_\pi^2A^{CD}-\triangle_\pi^{CD}
-{\triangle_\sigma}\per
\eea
It would be instructive to carry out this program with the available
information on the $\pi\pi$ amplitude. Finally, we comment on the
strangeness content of the pion,
\bea
y_\pi=\frac{2\langle\pi^0|\bar{s}s|\pi^0\rangle}{\langle
\pi^0|\bar{u}u+\bar{d}d|\pi^0\rangle}\per
\eea
{}From  the expressions  for the meson masses
to one loop ChPT \cite{glan}, one finds that $y_\pi$ is of the order of a
few percent. It is so small, because it is chirally suppressed,
$y_\pi=O(M_\pi^2)$.

For illustration, we note that - at one-loop accuracy \cite{glan} -
$\sigma_\pi\sim 69\MeV$.

\setcounter{equation}{0}
\section{Sigma-term in $\pi N$ scattering}
The analysis goes through in an analogous manner. The low-energy
 theorem reads in this case \cite{bpp}
\bea
F_\pi^2\bar{D}^{CD}_{\pi N}=\sigma_{\pi N}(2M_\pi^2)+\triangle_{\pi N}^{CD}\co
\eea
where $\bar{D}^{CD}_{\pi N}$ denotes the isospin even $D$-amplitude of
pion-nucleon scattering with pseudo-vector Born term subtracted. The
 quantity $\sigma_{\pi N}(t)$ is called the ``scalar form factor
of the proton'',
\bea
\bar{u}'u\sigma_{\pi N}(t)=\langle
 p'|\hat{m}(\bar{u}u+\bar{d}d)|p\rangle\sem t=(p'-p)^2\co
\eea
where $|p\rangle$ denotes a one-proton state.
 The sigma-term and the strangeness content of the proton are
\bea
\sigma_{\pi N}=\sigma_{\pi N}(0)\co
y_N=\frac{2\langle p|\bar{s}s|p\rangle}{\langle
  p|\bar{u}u+\bar{d}d| p\rangle}\per
\eea
The main point is that one can again use ChPT to determine
 $\triangle_{\pi N}^{CD}$: Whereas both the amplitude and the scalar
 form factor contain strong infrared
 singularities  \cite{meissner,bl,becherth}, these cancel in the
 present case completely up to and including terms of order
 $p^4$ \cite{meissner}, $\triangle_{\pi N}^{CD}=cM_\pi^4 +O(M_\pi^5)$,
 where the constant $c$ is quark mass independent. Numerically,
 the correction is very small \cite{meissner},
\bea
 \triangle_{\pi N}^{CD}\sim 2\MeV\per
\eea
There are two coherent phase shift analyses available, KH80 and
 KA85. These then lead to a coherent analysis of the
sigma-term \cite{gls},
\bea
\sigma_{\pi N}\simeq 45\MeV \scs y_N\simeq 0.2\co\nonumber\\
\sigma_{\pi N}(2M_\pi^2)-\sigma_{\pi N}(0)\simeq 15\MeV\per
\eea
This value for the sigma-term  has recently been confirmed - using  a different
 approach - by B\"uttiker and Mei\ss ner \cite{buettiker}.

The phase shifts from  KH80 and KA85 are essentially
based on data acquired in the 70's -
new data are included in the VPI/GW partial wave analyses. The most
recent version is SP00 extending to 2.1 GeV above which KA84
amplitudes are employed. We plan to report \cite{gs}  on the
impact of e.g. SP00 in an analysis similar to the one performed in
 Ref. \cite{gls}.

\setcounter{equation}{0}
\section{Kaon-nucleon sigma-terms }
The investigation of the kaon-nucleon  channels is much more
involved, because
\begin{itemize}
\item
there are less data,
\item there are open channels below threshold,
\item
there is a resonance just below threshold in $\bar{K}N\rightarrow \bar{K}N$.
\end{itemize}

We refer the  reader to Olin's contribution \cite{olin} for a
detailed discussion of these issues, see also \cite{reviewkn}.

There are two $\sigma$-terms in this case \cite{sigmakn}, and we denote
them by
\bea
\sigma_{K N}^u&=&\frac{\hat{m}+m_s}{4m_p}\langle p|\bar{u}u
+\bar{s}s|p\rangle\co\nonumber\\
\sigma_{K N}^d&=&\frac{\hat{m}+m_s}{4m_p}\langle p|\bar{d}d
+\bar{s}s|p\rangle\co
\eea
where $|p\rangle$ again denotes a one-proton state, and where $m_p$
stands for the proton mass.

Although much work has already been
 performed \cite{reviewkn,sigmakn,reya}, we believe that this  is  a
field where many things remain to be done - we present some of the
topics in the following agenda.

\begin{center}
{\bf Agenda }
\end{center}
\begin{itemize}
\item
Establish the analogue of the low-energy theorem (\ref{eq22}),
without the use of formal manipulations with
$T$-products, e.g. by working with the  generating
functional \cite{gss,bl}. Of course, $SU(3)$ breaking effects must be taken
into account.
In the following, we take it that the relation is
\bea
F_K^2D_{K N}(t,\nu) =\sigma_{K N}^u(t)+q'_\mu q_\nu r^{\mu\nu}_{KN}\co
\eea
where $D_{KN}$ denotes the crossing symmetric amplitude,
 and where the remainder $r^{\mu\nu}_{KN}$ again has the same singularity
structure as the amplitude itself.
\item
At the CD point $t=2M_K^2,\nu=0$, it then follows that
\bea
F_K^2D_{K N}^{CD}=\sigma_{K N}^u(2M_K^2)+\triangle_{KN}^{CD}\per
\eea
\begin{itemize}
\item[-]evaluate $D_{KN}^{CD}$ from data, e.g., relate it to scattering
lengths \cite{reya}. Prepare to make  use of the precise data that are
expected from the DEAR \cite{guaraldo} experiment. Note that the distance from
threshold to the CD point is considerably larger here than in $\pi\pi$
or $\pi N$ scattering.
\item[-]
evaluate the remainder $\triangle_{KN}^{CD}$ in ChPT. Determine its infra\-red
singu\-lari\-ties. Check whether the correction is small compared
  to $\sigma_{KN}^u$.
\item[-]
evaluate $\sigma_{KN}^{u}(2M_K^2)-\sigma_{KN}^u(0)$ from dispersion
relations.
\end{itemize}
\item
Make use of chiral symmetry and ChPT whenever possible.
\end{itemize}
Why is all this  interesting? One of the reasons is the fact that the
sigma-term $\sigma_{KN}^u$ can be related to the pion-nucleon
sigma-term plus a remainder that is expected to be small,
\bea\label{eq44}
\sigma_{KN}^u=({1+y_N})\frac{\hat{m}+m_s}{\hat{4m}}
\sigma_{\pi N} +\sigma_{KN}^{I=1}\co
\eea
where
\bea
\sigma_{KN}^{I=1}&=&
\frac{\hat{m}+m_s}{8m_p}\langle p|\bar{u}u-\bar{d}d|p\rangle
              \co
\eea
and where isospin refers to the $t$-channel. The isospin zero part is
\bea
\sigma_{KN}^{I=0}&=&
\frac{\hat{m}+m_s}{8m_p}\langle p|\bar{u}u +\bar{d}d
+2\bar{s}s|p\rangle\per
\eea
The isospin one part is expected to be small,
\bea
\sigma_{KN}^{I=1}\sim\frac{m_s+\hat{m}}{m_s-\hat{m}}
\frac{m_\Xi^2-m_\Sigma^2}{8m_p}\sim 50\MeV\per
\eea
{}From Eq. (\ref{eq44}),  we con\-clude that one can de\-ter\-mine  $y_N$, by
measuring $\sigma_{KN}^u$, $\sigma_{\pi N}$ and by estimating
 the isovector part $\sigma_{KN}^{I=1}$.

\setcounter{equation}{0}
\section{Off-shell methods}
There are also Ward identities that relate the off-shell amplitudes to
 the sigma-term directly. In case of the $\pi\pi$ scattering
amplitude, the relevant relation is
\bea
F_\pi^2A(t,\nu;q'^2,q^2)=M_\pi^{-2}(q'^2+q^2-M_\pi^2)\Gamma_\pi(t)
+q'^\mu q^\nu \bar{r}_{\mu\nu}\co
\eea
where the off-shell amplitude is the one obtained by using the
 divergence of the axial current as the interpolating field for the pion.
{}From this equation follow several exact relations, like
\bea
A(0,0;0,0)&=&-{2M_\pi F_\pi^{-2}\sigma_\pi}\co\nonumber\\
A(M_\pi^2,0;0,M_\pi^2)&=&0\co\nonumber\\
A(M_\pi^2,0;M_\pi^2,0)&=&0\per
\eea
The last two relations display the Adler zeros of the amplitude.  One may
now try to relate the off-shell point $t=\nu=q'^2=q^2=0$ to the
on-shell amplitude in the physical region, by use of the Adler zeros as
constraints on the interpolation procedure. While the idea is
beautiful, the beauty has its price: It is very
difficult to control the approximations made in this extrapolation.

\setcounter{equation}{0}
\section{Summary and conclusions}

1. Data, ChPT  and dispersion relations allow one to pin down

\vskip.5cm

\begin{tabular}{ll}
$\sigma_{\pi \pi}$&{\mbox very precisely},\\
$\sigma_{\pi N}$& {\mbox less precisely},\\
$\sigma_{KN}^{u,d}$&{\mbox barely until now.}
\end{tabular}

\vskip.5cm
\noindent
2. DEAR may improve the situation as far as the kaon sigma-term is
   concerned, provided that one succeeds to relate this quantity in a
   reliable manner to the scattering lengths, such that the ones
   determined by the DEAR experiment \cite{guaraldo} enter in a dominant manner.

\noindent
3. This issue is a challenge for theoretical physicists, in
   particular, for the chiral symmetry framework.

\noindent
4. As an intermediate step, it would be instructive  to perform
the  analysis  for
$\pi K\rightarrow \pi K$ \cite{roessl}.

\subsection*{Acknowledgements}
It is a pleasure to thank the organizers for the very informative week
that we have spent in Frascati.
This work  has been supported in part by the Swiss National Science
Foundation, and by TMR, BBW-Contract No. 97.0131  and  EC-Contract
No. ERBFMRX-CT980169 (EURODA$\Phi$NE).

\end{document}